\documentclass[aps,prd,groupedaddress,preprint]{revtex4}
\usepackage{amsmath, amsfonts, graphicx}

\begin{document}


\title{The Supersymmetric Leptophilic Higgs Model}

\author{Gardner Marshall}\email[]{grmarshall@wm.edu}
\author{Marc Sher}\email[]{mtsher@wm.edu}
\affiliation{Particle Theory Group, Department of Physics, College
of William and Mary, Williamsburg, VA 23187, USA}

\date{\today}

\begin{abstract}

In the leptophilic model, one Higgs doublet couples to quarks and
another couples to leptons. We study the supersymmetric version of
this model, concentrating on the tightly constrained Higgs sector,
which has four doublets. Constraints from perturbativity, unitarity,
and LEP bounds are considered. It is found that the lightest Higgs,
$h$, can have a mass well below $114$ GeV, and for masses
 below $100$ GeV will have a substantially enhanced branching
ratio into $\tau$ pairs. For this region of parameter space,
traditional production mechanisms (Higgs-strahlung, W fusion and
gluon fusion) are suppressed, but it may be produced in the decay of
heavier particles. The second lightest Higgs has a mass of
approximately $110$ GeV for virtually all of parameter space, with
Standard Model couplings, and thus an increase of a few GeV in the
current lower bound on the Standard Model Higgs mass would rule out
the model. The two heavier Higgs are both gauge-phobic, one decays
almost entirely into $b\bar{b}$ and can be produced via gluon fusion
while the other decays almost entirely into $\tau^+\tau^-$ but can't
be easily produced.

\end{abstract}



\maketitle

\section{Introduction}
\label{sec:Intro}

The main purpose of the Large Hadron Collider (LHC) is the study of
the mechanism of electroweak symmetry breaking
(EWSB). One of the simplest and most studied extensions of the
Standard Model is the Two Higgs Doublet Model (2HDM), in which two
scalar doublets are jointly responsible for electroweak symmetry
breaking and fermion mass acquisition \cite{hhg, Sher:1988mj}. This
model has a very rich phenomenology, including charged scalars and
pseudoscalars. Among the earliest motivations for the 2HDM is its
additional CP violation relative to the Standard Model \cite{
Lee:1973iz, Weinberg:1976hu, Branco:1985aq, Liu:1987ng,
Weinberg:1990me, Wu:1994ja, Accomando:2006ga}, which can provide an
additional source of baryogenisis and the relative abundance of
matter to antimatter in the universe \cite{Riotto:1999yt,
Dine:2003ax}. It was also motivated by the fact that supersymmetric
models and models with a Peccei-Quinn symmetry \cite{Peccei:1977hh}
will always require a minimum of two Higgs doublets.

In order to avoid unobserved tree-level flavor changing neutral
currents (FCNCs), all fermions with the same quantum numbers (and
which are thus capable of mixing) must couple to the same Higgs
multiplet. The Glashow-Weinberg theorem \cite{Glashow:1976nt} states
that a necessary and sufficient condition for the absence of FNCNs
at tree-level is that all fermions of a given charge and helicity
transform according to the same irreducible representation of
$SU(2)$, correspond to the same eigenvalue of $T_3$, and that a
basis exists in which they receive their contributions in the mass
matrix from a single source. In the 2HDM, this is due to the
introduction of discrete or continuous symmetries. Generally one may
either take both up and down type quarks to couple to the same
doublet or have each couple to its own doublet. It is usually
assumed that the leptons couple to the same doublet as the down type
quarks, in which case the former scenario describes the Type I 2HDM
while the latter describes the Type II 2HDM. Such couplings can be
enforced by imposing a suitable $\mathbb{Z}_{2}$ symmetry, which may
simply be imposed \textit{ad hoc} or which may arise as a subgroup
of a continuous symmetry (as in Peccei-Quinn or supersymmetric
models).

Despite the traditional convention that leptons couple to the same
doublet as the down type quarks, there is no a priori reason why
this must be the case. An alternative possibility is that both the
up and down type quarks couple to one doublet while the leptons
couple to the remaining doublet. While the traditional 2HDMs have
received a great deal of attention, relatively little work has been
done in investigating this alternative possibility. Those who have
focused on this model \cite{Su:2009fz, Logan:2009uf, Goh:2009wg,
Cao:2009as, Aoki:2009ha} have referred to it by several names, our
selection of which is the Leptophilic Two Higgs Doublet Model
(L2HDM). As noted by Su and Thomas \cite{Su:2009fz}, the
consequences of a L2HDM could drastically alter the possible
detection channels for a light Higgs at the LHC, so it is important
that it be considered as incoming data begins to arrive.
Furthermore, the possibility of  substantially enhanced leptonic
couplings (which can only occur in leptophilic models) may shed some
insight into explaining recent experimental results from PAMELA,
Fermi LAT, and H.E.S.S. \cite{Goh:2009wg}.

There also remain alternative possibilities. One can couple the
up-type quarks and leptons to one Higgs doublet and the down-type
quarks to the other (referred to as the ``flipped" model
\cite{Logan:2010ag}) or one can couple all of the charged fermions
to one doublet and the right-handed neutrino to another (referred to
as the ``neutrino-specific" model) \cite{Davidson:2009ha}. While
interesting in their own right, these models do not offer the
possibility of substantially enhanced leptonic couplings, and we
will not focus on them.

The most popular extension of the Standard Model is supersymmetry,
which can solve the hierarchy problem and which has a very tightly
constrained Higgs sector. Thus, one is led to consider the
supersymmetric versions of these alternative 2HDM models. Recently,
with McCaskey, we considered \cite{Marshall:2009bk} the
supersymmetric version of the ``neutrino-specific" model, and found
some remarkable signatures, including pentalepton and hexalepton
events with very high rates at the Tevatron and the LHC. In this
work, we extend the L2HDM to incorporate supersymmetry. The
resulting Supersymmetric Leptophilic Higgs Model (SLHM) leads to
exciting phenomenological prospects. In the scalar sector, the
strong constraints on the Higgs potential will substantially alter
the phenomenology of the lightest Higgs boson, since decays to
leptons can be substantially enhanced, and the decrease in the
coupling to the gauge bosons means that the current LEP bounds will
not apply, and much lighter Higgs bosons can be tolerated. In
addition, the supersymmetric partners to the leptons and the
leptonic Higgs doublet are influenced by the unusual Yukawa
structure. In the case of R-parity violation, the lightest
supersymmetric particle (LSP) could decay into leptons. Without
R-parity violation the LSP might annihilate into leptons
\cite{Goh:2009wg}. In this paper, we will focus on the scalar
sector, since the results may be testable in the very near future at
the Tevatron.

The layout of this paper is as follows. In Section \ref{sec:L2HDM}
we review the setup of the L2HDM. In Section \ref{sec:SLHM} we
introduce the SLHM and calculate the scalar mass matrices. In
Section \ref{sec:Constraints} we consider various constraints on the
model's parameter space by focusing on the neutral scalar sector. By
combining results from Yukawa coupling perturbativity
considerations, unitarity requirements, and direct searches for
Higgs bosons at LEP, we obtain severe restrictions on the model's
parameter space. In Section \ref{sec:Phenomenology} we discuss the
phenomenology of the lightest and next-to-lightest Higgs bosons at
the Tevatron and the LHC, and then in Section \ref{sec:Conclusion},
we conclude.

\section{The Leptophilic Two Higgs Doublet Model}
\label{sec:L2HDM}

The L2HDM contains two scalar $SU(2)_{L}$ doubles $\Phi_{q}$ and
$\Phi_{\ell}$. A discrete $\mathbb{Z}_{2}$ symmetry is imposed under
which $\Phi_{\ell} \rightarrow -\Phi_{\ell}$ and $e_{R_{i}}
\rightarrow -e_{R_{i}}$, but all other fields are invariant. The
resulting Yukawa lagrangian is given by \begin{equation}
\label{eq:L2HDM Yukawa} \mathcal{L}_{Y} = - \Big\{
Y_{ij}^{u}\overline{u}_{R_{i}}\widetilde{\Phi}_{q}^{\dag} \cdot
Q_{L_{j}} + Y_{ij}^{d}\overline{d}_{R_{i}}\Phi_{q}^{\dag} \cdot
Q_{L_{j}} + Y_{ij}^{\ell}\overline{e}_{R_{i}} \Phi_{\ell}^{\dag}
\cdot E_{L_{j}} + \textrm{h.c.} \Big\}, \end{equation} where
\[Q_{L_{i}} = \left( \begin{array}{c}
u_{L_{i}} \\
d_{L_{i}} \\
\end{array} \right), \hspace{.06 in} E_{L_{i}} = \left(
\begin{array}{c}
\nu_{L_{i}} \\
e_{L_{i}} \\
\end{array} \right), \hspace{.06 in} \textrm{and}
\hspace{.1 in} \Phi_{X} = \left( \begin{array}{c}
\phi_{X}^{+} \\
\frac{1}{\sqrt{2}}\big(v_{X} + \phi_{Xr}^{0} + i\phi_{Xi}^{0}\big) \\
\end{array} \right)\] for $X = q,\ell$ and $\widetilde{\Phi}_{q} =
i\sigma_{2}\Phi_{q}$. The Higgs sector potential is given by
\cite{Su:2009fz, Gunion:2002zf}
\begin{equation}
\begin{split} \label{eq:L2HDM Higgs Potential} V = {} &
m_{q}^{2}|\Phi_{q}|^{2} + m_{\ell}^{2}|\Phi_{\ell}|^{2} + \Big(m_{q
\ell}^{2} \Phi_{q}^{\dag}\Phi_{\ell} + \textrm{h.c.}\Big) +
\frac{\lambda_{1}}{2}|\Phi_{q}|^{4} +
\frac{\lambda_{2}}{2}|\Phi_{\ell}|^{4}
\\ & + \lambda_{3}|\Phi_{q}|^{2}|\Phi_{\ell}|^{2} +
\lambda_{4}|\Phi_{q}^{\dag}\Phi_{\ell}|^{2} +
\frac{\lambda_{5}}{2}\Big[\big(\Phi_{q}^{\dag}\Phi_{\ell}\big)^{2} +
\textrm{h.c.}\Big]. \end{split} \end{equation}

The physical scalars consist of two neutral scalars $h$ and $H$, a
pseudoscalar $\chi^{0}$, and a charged pair $H^{\pm}$. The other
three degrees of freedom are the Goldstone bosons $G^{\pm}$ and
$G^{0}$, which are eaten by the $W^{\pm}$ and $Z^{0}$ respectively.
If one defines the mixing angle $\tan\beta = v_{q}/v_{\ell}$, the
physical charged scalars can be expressed as \begin{equation}
\label{eq:L2HDM Charged Scalars} \left( \begin{array}{c}
G^{+} \\
H^{+} \\
\end{array} \right) = \left( \begin{array}{cc}
\cos\beta & \sin\beta \\
-\sin\beta & \cos\beta \\
\end{array}\right) \left( \begin{array}{c}
\Phi_{\ell}^{+} \\
\Phi_{q}^{+} \\
\end{array} \right). \end{equation} The physical neutral scalar
states are expressed in terms of the mixing angle $\tan\alpha$,
which can be solved for in terms of the entries of the neutral
scalar mass-squared matrix $\tan2\alpha = 2M_{12}^{2}/(M_{11}^{2} -
M_{22}^{2})$. One then finds the following relation \begin{equation}
\label{eq:L2HDM Neutral Scalars} \left(
\begin{array}{c}
H \\
h \\
\end{array} \right) = \sqrt{2} \left( \begin{array}{cc}
\cos\alpha & \sin\alpha \\
-\sin\alpha & \cos\alpha \\
\end{array}\right) \left( \begin{array}{c}
\phi_{\ell \hspace{.01 in} r}^{0} - v_{\ell} \\
\phi_{q r}^{0} - v_{q} \\
\end{array} \right). \end{equation} The vertex factors for the
couplings between the charged scalar and fermions are given by
\cite{Logan:2009uf} \begin{equation} \begin{split}
\label{eq:L2HDM Charged Scalar Quarks Coupling} H^{+}u_{i} d_{j} &
\rightarrow \left(\frac{ig \cot\beta}{2 \sqrt{2} M_{W}}\right)
V_{ij} \big[(m_{u_{i}} - m_{d_{j}}) - (m_{u_{i}} +
m_{d_{j}})\gamma_{5} \big], \\ H^{+}\nu_{i}e_{i} & \rightarrow
\left(\frac{ig \tan\beta}{2 \sqrt{2} M_{W}}\right) m_{e_{i}}
(1-\gamma_{5}).
\end{split} \end{equation} For large $\tan\beta$ the neutrino-lepton
coupling to $H^{+}$ is magnified while the quarks' coupling to
$H^{+}$ is diminished. The neutral scalar couplings to the charged
leptons will similarly be magnified. An interesting feature of the
model is that $\tan\beta$ can be much larger than in the
conventional 2HDMs without causing problems with perturbativity and
unitarity, since the Standard Model leptonic couplings are smaller
than the quark couplings.

\section{The Supersymmetric Leptophilic Higgs Model}
\label{sec:SLHM}

In this section we introduce the minimal leptophilic model required
to incorporate supersymmetry. A SLHM will require a minimum of four
Higgs doublets in order to achieve anomaly cancelation. Therefore,
we add to the MSSM two Higgs doublets $H_{0}$ and $H_{\ell}$ with
weak hypercharge assignments $+1/2$ and $-1/2$ respectively. The
four Higgs doublets along with their weak hypercharges are listed in
the table.

\begin{center}\begin{tabular}{c|cccc} $\Phi$ & $H_{u}$ & $H_{d}$ &
$H_{0}$ & $H_{\ell}$ \\ \hline $U_{Y}(1)$ & $+1/2$ & $-1/2$ & $+1/2$
& $-1/2$ \\ \end{tabular} \end{center}

The scalar doublets $H_{u}$ and $H_{d}$ are responsible for giving
mass to the up and down quarks respectively. We refer to these
doublets as the quark friendly doublets. Of the new doublets, the
lepton friendly doublet $H_{\ell}$ gives mass to the leptons, while
the remaining inert doublet $H_{0}$ does not couple to quarks or
leptons. This Yukawa structure is enforced by a discrete
$\mathbb{Z}_{2}$ symmetry, under which the superfields $E, H_{0}$,
and $H_{\ell}$ transform as $X \rightarrow -X$ while all other
fields remain unchanged. The most general superpotential respecting
R-parity, gauge symmetry, and the $\mathbb{Z}_{2}$ symmetry is
\begin{equation} \label{eq:Superpotential} W = y_{u}UQH_{u} -
y_{d}DQH_{d} - y_{\ell}ELH_{\ell} + \widetilde{\mu}_{1}H_{u}H_{d} +
\widetilde{\mu}_{2}H_{0}H_{\ell} \hspace{.01 in}. \end{equation}

The $\mathbb{Z}_{2}$ symmetry is softly broken by the terms
$(\mu_{3}^{2} H_{u} H_{\ell} + \mu_{4}^{2} H_{0} H_{d} +
\textrm{h.c.} )$ contained in the Higgs sector soft SUSY breaking
potential $V_{\textrm{Soft}}$ given by
\[V_{\textrm{Soft}} = \mu_{u}^{2} |H_{u}|^{2} + \mu_{d}^{2}
|H_{d}|^{2} + \mu_{0}^{2} |H_{0}|^{2} + \mu_{\ell}^{2}
|H_{\ell}|^{2} + \Big( \mu_{1}^{2} H_{u} H_{d} + \mu_{2}^{2}
H_{0}H_{\ell} + \mu_{3}^{2} H_{u} H_{\ell} + \mu_{4}^{2} H_{0} H_{d}
+ \textrm{h.c.} \Big). \] The Higgs sector potential is given by the
sum of the F-terms, D-terms, and $V_{\textrm{Soft}}$ respectively
\[V = \sum_{i=1}^{k} \left| \frac{\partial W}{\partial H_{i}}
\right|^{2} + \frac{1}{2} \hspace{.02 in} \sum_{a} \left|
\sum_{i=1}^{k} g^{a} H^{\dag}_{i} T^{a} H_{i} \right|^{2} +
V_{\textrm{Soft}}.\] Expanding the above expression results in
\begin{align*} V = {} & m_{u}^{2} |H_{u}|^{2} + m_{d}^{2}
|H_{d}|^{2} + m_{0}^{2} |H_{0}|^{2} + m_{\ell}^{2} |H_{\ell}|^{2} +
\big( \mu_{1}^{2} H_{u} H_{d} + \mu_{2}^{2} H_{0}H_{\ell} +
\mu_{3}^{2} H_{u} H_{\ell} + \mu_{4}^{2} H_{0} H_{d} + \textrm{h.c.}
\big) \\ & + \frac{g_{1}^{2}}{8} \sum_{a} \left| H_{u}^{\dag}
\sigma^{a} H_{u} + H_{d}^{\dag} \sigma^{a} H_{d} + H_{0}^{\dag}
\sigma^{a} H_{0} + H_{\ell}^{\dag} \sigma^{a} H_{\ell} \right|^{2} +
\frac{g_{2}^{2}}{8} \Big| |H_{u}|^{2} - |H_{d}|^{2} + |H_{0}|^{2} -
|H_{\ell}|^{2} \Big|^{2}, \end{align*} where $m_{u}^{2} =
(|\widetilde{\mu}_{1}|^{2} + \mu_{u}^{2}), m_{d}^{2} =
(|\widetilde{\mu}_{1}|^{2} + \mu_{d}^{2}), m_{0}^{2} =
(|\widetilde{\mu}_{2}|^{2} + \mu_{0}^{2}), m_{\ell}^{2} =
(|\widetilde{\mu}_{2}|^{2} + \mu_{\ell}^{2})$, and $\sigma^{a}$ ($a
= 1,2,3$) are the Pauli matrices. To achieve spontaneous symmetry
breaking, the Higgs doublets acquire the following vacuum
expectation values (vevs): \begin{equation} \langle
H_{u} \rangle = \frac{1}{\sqrt{2}} \left( \begin{array}{c} 0 \\
v_{u} \\ \end{array} \right), \hspace{.1 in} \langle H_{d} \rangle =
\frac{1}{\sqrt{2}} \left( \begin{array}{c} v_{d} \\
0 \\ \end{array} \right), \langle H_{0} \rangle = \frac{1}{\sqrt{2}}
\left( \begin{array}{c} 0 \\ v_{0} \\ \end{array} \right),
\hspace{.1 in} \langle H_{\ell} \rangle = \frac{1}{\sqrt{2}} \left(
\begin{array}{c} v_{\ell} \\ 0 \\ \end{array} \right).
\end{equation} We define $v^{2} = v_{u}^{2} + v_{d}^{2} + v_{0}^{2}
+ v_{\ell}^{2}$ so that we have $v^{2} = 4M_{Z}^{2}/(g_{1}^{2} +
g_{2}^{2}) \approx (246 \ \textrm{GeV})^{2}$. Between the quark
friendly doublets we define the mixing angle $\tan\beta =
v_{u}/v_{d}$ while between the lepton friendly and inert doublets we
define the mixing angle $\tan\beta_{\ell} = v_{0}/v_{\ell}$. We also
define $\tan\alpha = v_{q}/v_{L}$, where $v_{q}^{2} = v_{u}^{2} +
v_{d}^{2}$ and $v_{L}^{2} = v_{0}^{2} + v_{\ell}^{2}$. These
definitions allow us to express the individual vevs in terms of the
Standard Model vev and the three mixing angles $\alpha, \beta$, and
$\beta_{\ell}$ \begin{equation} \label{eq:vev breakdowns} v_{u} = v
\sin\alpha \sin\beta, \hspace{.2 in} v_{d} = v \sin\alpha \cos\beta,
\hspace{.2 in} v_{0} = v \cos\alpha \sin\beta_{\ell}, \hspace{.2 in}
v_{\ell} = v \cos\alpha \cos\beta_{\ell}. \end{equation}

Each of the four complex Higgs doublets contains four real degrees
of freedom, so there are a total of sixteen degrees of freedom.
Three of these are eaten to give mass to the $W^{\pm}$ and $Z^{0}$,
while those remaining result in a scalar mass spectrum that includes
four neutral scalars, three pseudoscalars, and three charged pairs.
From the scalar potential above, the mass matrices can be
calculated. We parameterize them in terms of the gauge boson masses
and the three mixing angles appearing in equation \ref{eq:vev
breakdowns}.

The neutral scalar mass matrix is $M_{N}^{2} =$
\[\left( \begin{array}{cccc} M_{1}^{2} & -\frac{1}{2} M_{Z}^{2}
s^{2}_{\alpha} s_{2\beta} - \mu_{1}^{2} & \frac{1}{2}M_{Z}^{2}
s_{2\alpha} s_{\beta} s_{\beta_{\ell}} & -\frac{1}{2} M_{Z}^{2}
s_{2\alpha} s_{\beta} c_{\beta_{\ell}} -\mu_{3}^{2} \\ -\frac{1}{2}
M_{Z}^{2} s^{2}_{\alpha} s_{2\beta} - \mu_{1}^{2} & M_{2}^{2} &
-\frac{1}{2}M_{Z}^{2} s_{2\alpha} c_{\beta} s_{\beta_{\ell}} -
\mu_{4}^{2} & \frac{1}{2}M_{Z}^{2} s_{2\alpha} c_{\beta}
c_{\beta_{\ell}} \\ \frac{1}{2}M_{Z}^{2} s_{2\alpha} s_{\beta}
s_{\beta_{\ell}} & -\frac{1}{2}M_{Z}^{2} s_{2\alpha} c_{\beta}
s_{\beta_{\ell}} - \mu_{4}^{2} & M_{3}^{2} & -\frac{1}{2} M_{Z}^{2}
c^{2}_{\alpha} s_{2\beta_{\ell}} - \mu_{2}^{2} \\ -\frac{1}{2}
M_{Z}^{2} s_{2\alpha} s_{\beta} c_{\beta_{\ell}} -\mu_{3}^{2} &
\frac{1}{2} M_{Z}^{2}s_{2\alpha} c_{\beta} c_{\beta_{\ell}} &
-\frac{1}{2} M_{Z}^{2} c^{2}_{\alpha} s_{2\beta_{\ell}} -
\mu_{2}^{2} & M_{4}^{2} \\ \end{array} \right)\] where $s_{x}$ and $
c_{x}$ are shorthand for $\sin x$ and $\cos x$ respectively, and the
diagonal terms are given by \[\begin{array}{ccccccc} M_{1}^{2} & = &
M_{Z}^{2} \sin^{2}\alpha \sin^{2}\beta + \lambda_{1}, & \hspace{.1
in} & \lambda_{1} & = & \mu_{1}^{2} \cot\beta + \mu_{3}^{2}
\cot\alpha \left( \frac{\cos\beta_{\ell}}{\sin \beta} \right), \\
M_{2}^{2} & = & M_{Z}^{2} \sin^{2}\alpha \cos^{2}\beta +
\lambda_{2}, & \hspace{.1 in} & \lambda_{2} & = & \mu_{1}^{2}
\tan\beta + \mu_{4}^{2} \cot\alpha \left( \frac{\sin\beta_{\ell}}{
\cos\beta} \right), \\ M_{3}^{2} & = & M_{Z}^{2} \cos^{2}\alpha
\sin^{2}\beta_{\ell} + \lambda_{3}, & \hspace{.1 in} & \lambda_{3} &
= & \mu_{2}^{2} \cot\beta_{\ell} + \mu_{4}^{2} \tan\alpha \left(
\frac{\cos\beta}{\sin\beta_{\ell}} \right), \\ M_{4}^{2} & = &
M_{Z}^{2} \cos^{2}\alpha \cos^{2} \beta_{\ell} + \lambda_{4}, &
\hspace{.1 in} & \lambda_{4} & = & \mu_{2}^{2} \tan\beta_{\ell} +
\mu_{3}^{2} \tan\alpha \left( \frac{\sin\beta}{\cos\beta_{\ell}}
\right). \end{array}\] The pseudoscalar mass matrix is
\begin{equation} \label{eq:Pseudoscalar Mass Matrix} M_{A}^{2} =
\left( \begin{array}{cccc}
\lambda_{1} & \mu_{1}^{2} & 0 & \mu_{3}^{2} \\
\mu_{1}^{2} & \lambda_{2} & \mu_{4}^{2} & 0 \\
0 & \mu_{4}^{2} & \lambda_{3} & \mu_{2}^{2} \\
\mu_{3}^{2} & 0 & \mu_{2}^{2} & \lambda_{4} \\
\end{array} \right). \end{equation}  The charged
scalar mass matrix is \begin{equation} \label{eq:Charged Scalar Mass
Matrices} M_{H^{\pm}}^{2} = M_{A}^{2} + \Delta M^{2}, \end{equation}
where \[\Delta M^{2} = M_{W}^{2} \left( \begin{array}{cccc}
s^{2}_{\alpha} c^{2}_{\beta} + c^{2}_{\alpha}c_{2\beta_{\ell}} &
\frac{1}{2} s^{2}_{\alpha} s_{2\beta} & \frac{1}{2} s_{2\alpha}
s_{\beta} s_{\beta_{\ell}} & \frac{1}{2} s_{2\alpha} s_{\beta}
c_{\beta_{\ell}} \\ \frac{1}{2} s^{2}_{\alpha} s_{2\beta} &
s^{2}_{\alpha} s^{2}_{\beta} - c^{2}_{\alpha} c_{2\beta_{\ell}} &
\frac{1}{2} s_{2\alpha} c_{\beta} s_{\beta_{\ell}} & \frac{1}{2}
s_{2\alpha} c_{\beta}c_{\beta_{\ell}}
\\ \frac{1}{2} s_{2\alpha} s_{\beta} s_{\beta_{\ell}} & \frac{1}{2}
s_{2\alpha} c_{\beta} s_{\beta_{\ell}} & c^{2}_{\alpha}
c^{2}_{\beta_{\ell}} + s^{2}_{\alpha} c_{2\beta} & \frac{1}{2}
c^{2}_{\alpha} s_{2\beta_{\ell}} \\ \frac{1}{2} s_{2\alpha}
s_{\beta} c_{\beta_{\ell}} & \frac{1}{2} s_{2\alpha} c_{\beta}
c_{\beta_{\ell}} & \frac{1}{2} c^{2}_{\alpha} s_{2\beta_{\ell}} &
c^{2}_{\alpha} s^{2}_{\beta_{\ell}} - s^{2}_{\alpha} c_{2\beta} \\
\end{array} \right).\]

In Section 3.3 of \cite{Gupta:2009wn} Gupta and Wells outline a
procedure for obtaining an upper bound on the tree-level mass of the
lightest neutral scalar, $h$, in the limit of large SUSY breaking
masses (as compared to the $Z$-mass). The procedure consists of
transforming the mass matrices into the so called ``Runge basis," in
which one doublet obtains all of the vev while the others are
orthogonal to one another. Details on the Runge basis can be found
in \cite{Wells:2009kq}. In this basis all but one diagonal entry of
the neutral scalar mass matrix grow large in the limit of large SUSY
breaking masses. This entry acts as an upper bound on $M_{h}^{2}$
since, for a positive definite matrix, the smallest eigenvalue is
bounded above by the smallest diagonal entry. Their result holds in
our case as well and results in the inequality \begin{equation}
\label{eq:Gupta Bound} M_{h} \leq M_{Z}|\sin^{2}\alpha \cos2\beta +
\cos^{2}\alpha \cos2\beta_{\ell}|.
\end{equation}

Leading order radiative corrections to the Higgs masses will be
important in constraining parameter space. As usual, the dominant
contributions come from top quark loops, governed by the top quark
Yukawa coupling. In this section we have written the neutral scalar
mass matrix, $M^{2}_{N}$, in the $\{u,d,0,\ell\}$ basis. Hence the
1-1 entry receives a correction from top quark loop diagrams given
by \begin{equation} \label{eq:Radiative Correction To M11} \Delta
M_{11}^{2} = \frac{3 \alpha}{\pi} \left(\frac{m_{t}^{4}
}{M_{Z}^{2}}\right) \frac{\ln\big(m_{ \tilde{t}}^{2}
/m_{t}^{2}\big)}{\sin^{2}2\theta_{W} \sin^{2}\alpha\sin^{2}\beta},
\end{equation} where $m_{\tilde{t}}$ is the stop squark mass, which
we take to be $\sim 1$ TeV. In addition to top quark loop
corrections, other corrections are potentially significant because
of the possibility of very large values for $\tan\beta$ and
$\tan\beta_{\ell}$. We therefore also consider the leading
correction to the 2-2 and 4-4 entries of $M^{2}_{N}$, which come
from bottom quark loop diagrams and a tau loop diagram respectively.
The 3-3 entry receives no correction since the inert doublet,
$H_{0}$, does not couple to quarks or leptons.    There are other sub-leading-log corrections to the masses, and these can contribute $5-10$ GeV to the masses (see Ref. \cite{Djouadi:2005gj} for a detailed discussion).

\section{Constraints on The Supersymmetric Leptophilic Higgs Model}
\label{sec:Constraints}

In this section we outline the main constraints that limit the
viable parameter space of the SLHM. The free parameters arising from
the scalar sector consist of the four couplings $\mu_{1}^{2}$,
$\mu_{2}^{2}$, $\mu_{3}^{2},$ and $\mu_{4}^{2}$, which mix pairs of
Higgs doublets in the scalar potential, as well as the three mixing
angles $\tan\alpha, \tan\beta,$ and $\tan\beta_{\ell}$, which appear
in equation \ref{eq:vev breakdowns}. The constraints arising from
the charged scalar sector are similar to those of the L2HDM, which
is studied in \cite{Logan:2009uf}. Our interest therefore lies in
the neutral sector. We find that LEP data and other constraints
severely restrict the size of the allowable parameter space, but
leave enough room to comfortably fit the model a lightest
neutral scalar mass substantially less than $110$ GeV.

\subsection{Yukawa Coupling Perturbativity}
\label{sec:Yukawa Perturbativity}

The first constraints come from requiring that the Yukawa couplings
remain perturbative. By demanding that each Yukawa coupling remains
smaller than $4\pi$ we obtain the following three inequalities
\begin{align} \label{eq:Perturbativity Constraints} \begin{split} &
\bigg(1 + \frac{1}{\tan^{2}\alpha}\bigg) \bigg(1 +
\frac{1}{\tan^{2}\beta} \bigg) \hspace{.05 in} < \hspace{.05 in}
\frac{8 \pi^{2} v^{2}}{m_{t}^{2}} \approx 13^{2}, \\ & \bigg(1 +
\frac{1}{\tan^{2} \alpha}\bigg) \bigg(1 + \tan^{2}\beta \bigg)
\hspace{.05 in} < \hspace{.05 in} \frac{8 \pi^{2} v^{2}}{m_{b}^{2}}
\approx 520^{2}, \\ & \bigg(1+\tan^{2}\alpha\bigg) \bigg(1 +
\tan^{2}\beta_{\ell} \bigg) \hspace{.05 in} < \hspace{.05 in}
\frac{8 \pi^{2} v^{2}}{m_{\tau}^{2}} \approx 1235^{2}. \end{split}
\end{align} One can see that the top quark Yukawa coupling becomes
non-perturbative for small values of $\tan\alpha$ or $\tan\beta$
while the bottom quark Yukawa coupling does so for small values of
$\tan\alpha$ or large values of $\tan\beta$. In addition, the tau
Yukawa coupling becomes non-perturbative for large values of
$\tan\alpha$ or $\tan\beta_{\ell} \hspace{.02 in}$.

\subsection{Tree Level Unitarity}
\label{sec:Unitarity}

Requiring perturbative unitarity of fermion anti-fermion scattering
places upper bounds on the fermion masses. The unitarity condition
that must be satisfied is $|\Re(a_{J})| \leq 1/2$, where $a_{J}$ is
the $J$th partial wave amplitude in the partial wave expansion of
the fermion anti-fermion scattering amplitude. The scattering we
consider occurs by the exchange of a Higgs boson. We obtain bounds
from imposing the unitarity condition on the $J = 0$ partial wave
amplitude, which is calculated from a sum over s- and t-channel
helicity amplitudes in the high energy limit. The procedure is
described in detail in \cite{Dawson:2010jx}, where contributions to
the partial wave amplitudes are provided for a general model. These
contributions depend on combinations of the vector and axial vector
Yukawa couplings. For the SLHM the resultant bounds are found to be
(see \cite{Dawson:2010jx} for a clear discussion) \begin{align}
\label{eq:Unitarity Constraints} \begin{split} & \frac{G_{F}
m_{t}^{2}}{4 \pi \sqrt{2}} < \sin^{2}\alpha \sin^{2}\beta, \\ &
\frac{G_{F} m_{b}^{2}}{4 \pi \sqrt{2}} < \sin^{2}\alpha
\cos^{2}\beta, \\ & \frac{G_{F} m_{\tau}^{2}}{4 \pi \sqrt{2}} <
\cos^{2}\alpha \cos^{2}\beta_{\ell}. \end{split} \end{align} Here we
have used the bounds obtained for third generation fermions as their
larger masses yield the most stringent results. The unitarity
constraint prevents very large values for $\tan\beta$, capping it at
around 300. Several combinations of $\tan\alpha$ and $\tan\beta$
values on the order of several tenths are also eliminated.

\subsection{The Anomalous Muon Magnetic Moment}
\label{sec:Muon Magnetic Moment}

As in the Standard Model, the magnetic moment of the muon receives a
contribution from the one-loop diagram formed by connecting the muon
lines on a muon-muon-photon vertex with a neutral Higgs boson. Only
the lightest neutral Higgs is relevant since the contribution goes
as the square of the ratio between the muon and Higgs masses. For
the SLHM the contribution is \begin{equation} \label{eq:Muon
Magnetic Moment} \Delta a_{\mu} = K^{2} \hspace{.01 in}
\frac{m_{\mu}^{2}}{8 \pi^{2} v^{2}} \int_{0}^{1} \frac{z^{2}
(2-z)}{z^{2} + x^{2}(1-z)} \hspace{.03 in} dz, \end{equation} where
$x = M_{h}/m_{\mu}$ and \[K^{2} = \frac{|U_{41}|^{2}}{\cos^{2}\alpha
\cos^{2}\beta_{\ell}}.\] If the Higgs mass, $M_{h}$, is assumed to
be the same in the SLHM and the Standard Model then the contribution
to the muon's magnetic moment from a light scalar in the SLHM is
simply its Standard Model value multiplied by $K^{2}$. The value of
$K^{2}$ however, remains $\lesssim 1$ across the entire spectrum of
parameter space, even for very large values of $\tan\alpha$ and
$\tan\beta_{\ell}$. A review on the anomalous muon magnetic moment
is given by \cite{Jegerlehner:2007xe} while current results and
uncertainties can be found in \cite{Bennett:2004pv, Bennett:2006fi}.
In our case the contribution is much too small to produce any
bounds.

In addition however, there is a two-loop Barr-Zee effect
\cite{Barr:1990vd}, which is generally more significant than the
one-loop contribution discussed above. The Barr-Zee effect occurs by
connecting an internal Higgs to an internal photon through a massive
fermion loop and is given by \cite{Cheung:2003pw, El Kaffas:2006nt}.
We consider such effects with third generation fermions in the SLHM
and find that the contribution to the muon magnetic moment is
\begin{equation} \label{eq:Muon Magnetic Moment II} \Delta a_{\mu} =
-\frac{\alpha \hspace{.02 in} m_{\mu}^{2} U_{41}}{4 \pi^{3} v^{2}
\cos\beta_{\ell}} \left\{ \frac{8 U_{11} f(x_{t})}{3 \sin2\alpha
\sin\beta} + \frac{2U_{21} f(x_{b})}{3 \sin2\alpha \cos\beta} +
\frac{U_{41} f(x_{\tau})}{ \cos^{2}\alpha \cos\beta_{\ell}}
\right\},\end{equation} where $x_{f} = m_{f}^{2}/M_{h}^{2}$ and the
function $f(x)$ is given by \[f(x) = \frac{x}{2} \int_{0}^{1}\frac{1
- 2z(1-z)}{z(1-z)-x} \ln\left[\frac{z(1-z)}{x}\right] dz.\] Though
the contribution from the tau loop diagram is suppressed by
$m_{\tau}^{2}/M_{h}^{2}$, it is enhanced for very large
$\tan\beta_{\ell}$. In following \cite{WahabElKaffas:2007xd} we
measure how well these contributions compare to experiment with the
quantity \[\chi_{a_{\mu}}^{2} = \left(\frac{\Delta
a_{\mu}^{\textrm{SLHM}}}{6.8 \times 10^{-10}}\right)^{2} ,\] where
$6.8 \times 10^{-10}$ is the theoretical uncertainty for $a_{\mu}$
in the Standard Model (used because it is larger than the
experimental uncertainty). The result is that, though larger than
the one-loop contributions, the two-loop Barr-Zee effect
contributions are still too small to provide significant constraints
on the parameter space.

\subsection{LEP Higgs Search Data}
\label{sec:LEP Data}

The largest source of constraints for the neutral sector of the SLHM
consists of LEP's failure to discover a neutral Higgs boson. If the
lightest neutral scalar's mass is too small, one would expect LEP to
have seen it, whereas for a mass $M_{h} > 114.4$ GeV, LEP data
becomes irrelevant and no bounds can be obtained
\cite{Amsler:2008zzb}. The production mechanism at LEP is the
Higgs-strahlung process $e^{+}e^{-} \rightarrow hZ$,  and thus if
the coupling, $g_{ZZh}$, between the lightest neutral scalar and
$Z$-pairs is sufficiently small, the scalar's non-discovery at LEP
can be explained \cite{TeixeiraDias:2008fc, Barate:2003sz,
Schael:2006cr,Achard:2003ty}.

In addition, there is an effect which suppresses the
sensitivity with which the experimental results may be applied to
constrain models beyond the Standard Model \cite{Abdallah:2004wy,
WahabElKaffas:2007xd}. Bounds from LEP were produced under the
assumption that the Higgs boson decays exclusively into $b \bar{b}$
pairs or exclusively into $\tau^{+} \tau^{-}$ pairs. LEP has
provided a bound on the quantity $\textrm{BR}(h \rightarrow X
\overline{X}) \xi^{2}$ for $X = b$ and $X = \tau$, where $\xi$ is
the ratio of the $ZZh$ coupling in a  model to that of
the Standard model i.e. $\xi = g_{ZZh} / g_{ZZh}^{SM}$. We find the
value of $\xi^{2}$ in the SLHM to be \begin{equation}
\label{eq:Value of Xi} \xi^{2} = \Big|U_{11} \sin\alpha \sin\beta +
U_{21} \sin\alpha \cos\beta + U_{31} \cos\alpha \sin\beta_{\ell} +
U_{41} \cos\alpha \cos\beta_{\ell} \ \Big|^{2}. \end{equation} We
will employ both of these bounds to exclude regions of parameter
space in the SLHM. Naively, one expects $\textrm{BR}(h \rightarrow b
\overline{b})$ to approach unity when $\tan\beta$ is large and
$\tan\alpha$, $\tan\beta_{\ell}$ are small since in that case the
down-type quark Yukawa couplings are doubly enhanced while the
lepton Yukawa couplings remains small. On the other hand, when
$\tan\alpha$ and $\tan\beta_{\ell}$ are large while $\tan\beta$ is
small, the lepton Yukawa couplings are enhanced and the down-type
quark Yukawa couplings remain small, resulting in an increase in the
branching ratio $\textrm{BR}(h \rightarrow \tau^{+} \tau^{-})$.

Since in the interesting region of parameter space, the $ZZh$ and
$WWh$ couplings are small, we can approximate the total decay width
as simply $\Gamma(h \rightarrow b \bar{b}) + \Gamma(h \rightarrow
\tau^{+} \tau^{-})$. The two branching ratios for the SLHM can
therefore be conveniently expressed as $\textrm{BR}(h \rightarrow b
\overline{b}) = 1/(1 + \kappa)$ and $\textrm{BR}(h \rightarrow
\tau^{+} \tau^{-}) = \kappa/(1 + \kappa)$, where $\kappa = \Gamma(h
\rightarrow \tau^{+} \tau^{-})/\Gamma(h \rightarrow b \bar{b})$. The
variable $\kappa$ is straightforward to calculate and is given by
\begin{equation} \label{eq:Kappa BRtau-BRb Ratio} \kappa =
\left(\frac{m_{\tau}^{2} }{3 m_{b}^{2}}\right) \tan^{2}\alpha
\frac{\cos^{2}\beta }{\cos^{2}\beta_{\ell}} \left|\frac{U_{41}}{
U_{21}}\right|^{2} \left(\frac{M_{h}^{2} - 4m_{\tau}^{2}}{M_{h}^{2}
- 4m_{b}^{2}}\right)^{3/2}, \end{equation} where the $U_{ij}$ are
entries of the $4 \times 4$ diagonalizing matrix defined by
$U^{\dag}M_{N}^{2}U = M_{\textrm{diag}}^{2}$.

\begin{figure}[h]
\includegraphics[scale=.9]{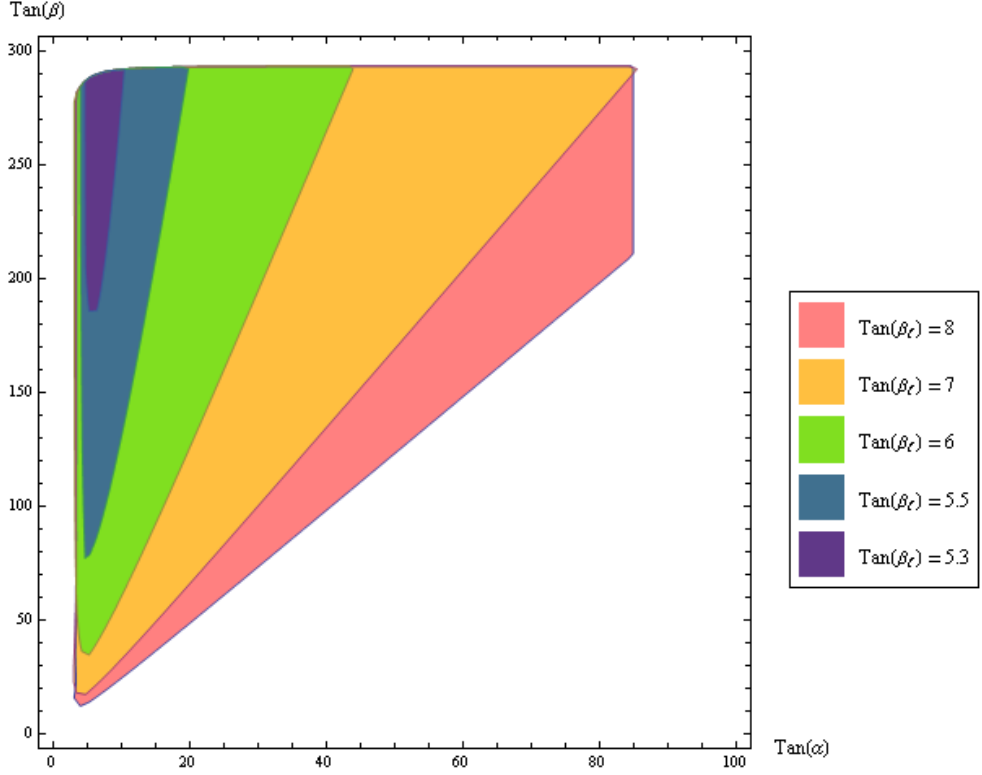}
\caption{ \small The colored regions illustrate the allowed points
in the $\tan\alpha$, $\tan\beta$, $\tan\beta_{\ell}$ parameter
space. Each region is a slice of constant $\tan\beta_{\ell}$ in the
$\tan\alpha \times \tan\beta$ plane. The values of $\mu_{1}$,
$\mu_{2}$, $\mu_{3}$, and $\mu_{4}$ are fixed at $200$, $250$,
$300$, and $100$ GeV respectively, but changing $\mu_{1}$ and/or
$\mu_{3}$ has relatively little effect. Increasing $\mu_{2}$ and/or
$\mu_{4}$ shrinks the above space. Increasing $\tan\beta_{\ell}$
enlarges the size of the allowed space quite rapidly until around
$\tan\beta_{\ell} \approx 8$, when the space stops enlarging and
begins to slowly shrink - this can be seen in Figure
\ref{fig:Plot2}.} \label{fig:Plot1}
\end{figure}

\begin{figure}[h]
\includegraphics[scale=.9]{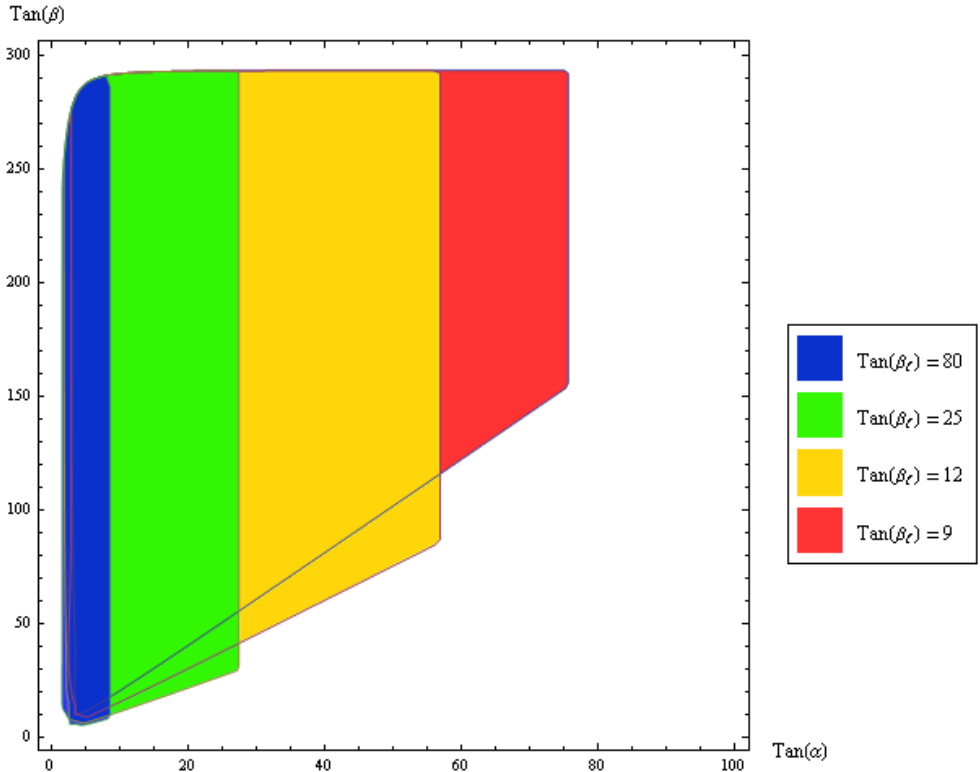}
\caption{ \small A continuation of figure \ref{fig:Plot1} for larger
values of $\tan\beta_{\ell}$. As $\tan\beta_{\ell}$ increases beyond
80, the space very slowly shrinks into an extremely thin sliver of
possible $\tan\alpha$ values centered near 2; it finally disappears
completely at $\tan\beta_{\ell} \approx 350$.} \label{fig:Plot2}
\end{figure}

We have numerically scanned through parameter space, calculating the
values of $\textrm{BR}(h \rightarrow b \overline{b}) \xi^{2},
\textrm{BR}(h \rightarrow \tau^{+} \tau^{-}) \xi^{2}$, and $M_{h}$
in the SLHM. Those points in parameter space for which either
$\textrm{BR}(h \rightarrow b \overline{b}) \xi^{2}$ or
$\textrm{BR}(h \rightarrow \tau^{+} \tau^{-}) \xi^{2}$ is greater
than its LEP bound at the corresponding value of $M_{h}$ are
excluded. By imposing these two LEP bounds as well as the
perturbativity requirements of Section \ref{sec:Yukawa
Perturbativity} and the unitarity requirements of Section
\ref{sec:Unitarity}, we are able to exclude substantial regions of
the model's parameter space. In Figures \ref{fig:Plot1} and
\ref{fig:Plot2} the allowed region of the three-dimensional
parameter space for the variables $\tan\alpha$, $\tan\beta$, and
$\tan \beta_{\ell}$ is shown. For these plots the values of
$\mu_{1}$, $\mu_{2}$, $\mu_{3}$, and $\mu_{4}$ have been fixed at
$200, 250, 300$, and $100$ GeV respectively. The plots depict
several sections of viable parameter space in the $\tan\alpha \times
\tan\beta$ plane, each being a slice of constant $\tan\beta_{\ell}$.
As $\tan\beta_{\ell}$ varies over its allowed range, one can see how
the sections grow in area, change shape, and eventually shrink back
away.

Though the values of $\mu_{1}$, $\mu_{2}$, $\mu_{3}$, and $\mu_{4}$
are fixed, the size and shape of the allowed parameter space remains
largely unchanged when $\mu_{1}$ and $\mu_{3}$ are allowed to vary
between $50$ and $1000$ GeV. Their values are consequentially
relatively unconstrained. Increasing the value of $\mu_{4}$ however,
has the effect of sharply cutting down on the size of the allowed
region of parameter space. So too does increasing $\mu_{2}$, though
to a slightly lesser degree. Merely increasing $\mu_{4}$ to $200$
GeV results in a drastically smaller allowed region than that shown
in Figure \ref{fig:Plot1} and completely eliminates the regions
corresponding to $\tan\beta_{\ell}$ values of $5.3$ and $5.5$. The
other regions are compressed so that $3 \lesssim \tan\alpha \lesssim
20$ and $50 \lesssim \tan\beta \lesssim 290$, while their overall
shape remains the same. Enlarging either $\mu_{2}$ or $\mu_{4}$
further rapidly shrinks the allowed space away until it vanishes
completely.

\begin{figure}[h]
\includegraphics[scale=1]{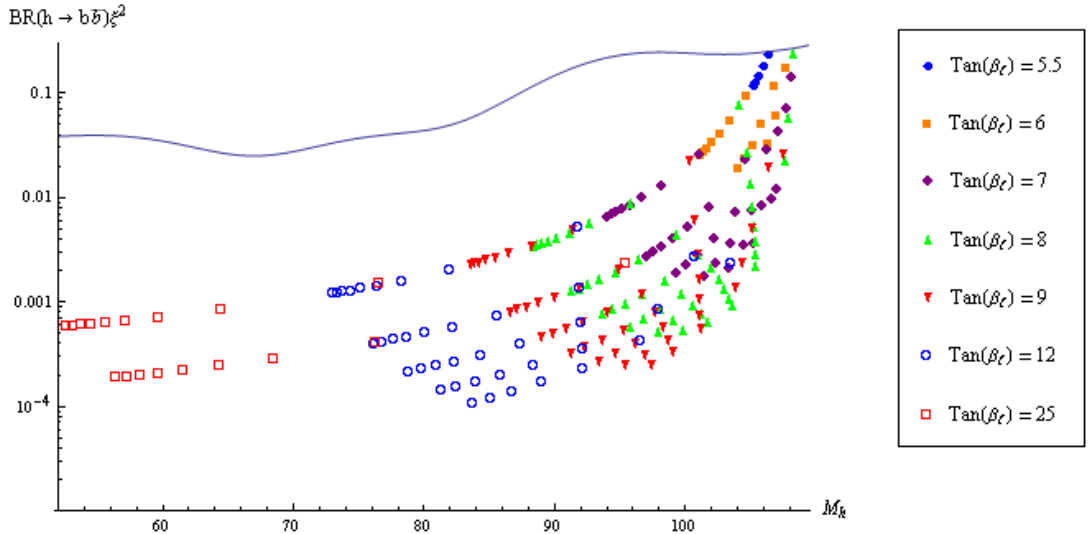}
\caption{ \small Various values of the quantity $\textrm{BR}(h
\rightarrow b\bar{b}) \xi^{2}$ plotted as a function of the lightest
neutral scalar mass $M_{h}$. The plotted values correspond to a
uniform sampling of points within the allowed regions of the
$\tan\alpha \times \tan\beta$ plane for the different values of
$\tan\beta_{ell}$ that are plotted in figures \ref{fig:Plot1} and
\ref{fig:Plot2}. The LEP bound of reference \cite{Barate:2003sz} is
shown in blue.} \label{fig:Plot3}
\end{figure}

\begin{figure}[h]
\includegraphics[scale=1]{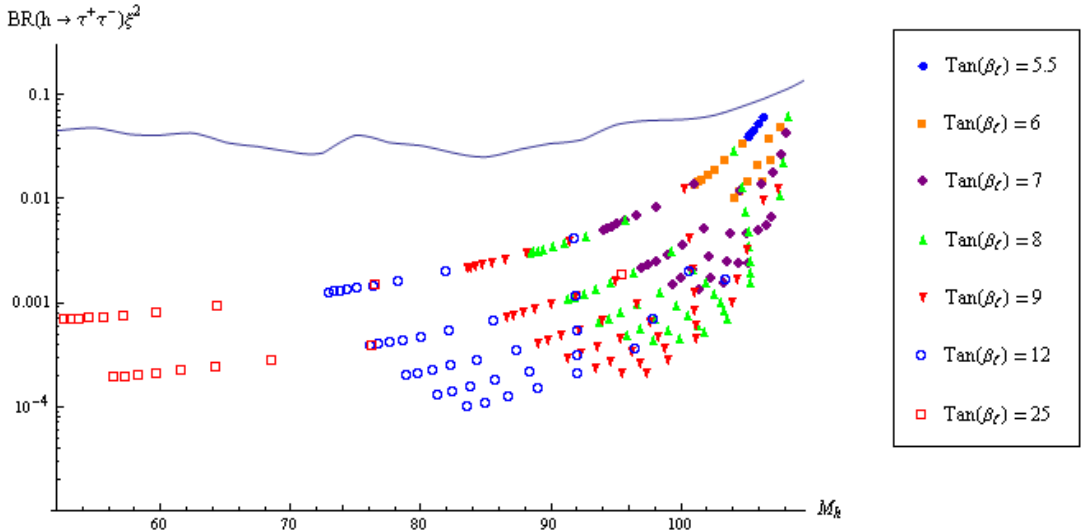} \caption{ \small Various values
of the quantity $\textrm{BR}(h \rightarrow \tau^{+}\tau^{-})
\xi^{2}$ plotted as a function of the lightest neutral scalar mass
$M_{h}$.} \label{fig:Plot4}
\end{figure}

Figure \ref{fig:Plot3} plots an assortment of possible
$\textrm{BR}(h \rightarrow b\bar{b}) \xi^{2}$ values as a function
of the lightest neutral scalar mass $M_{h}$. Each value plotted
corresponds to some point in the allowed region of parameter space.
The LEP curve is shown in blue. For very large values of
$\tan\beta_{\ell}$, the curves continue down to approximately $25$
GeV, with the value of $\textrm{BR}(h \rightarrow b\bar{b}) \xi^{2}$
becoming extremely small. We see that Higgs bosons below $114.4$ GeV
are certainly allowed, but below approximately $90$ GeV their
couplings to vector bosons become negligible, making detection
through vector boson fusion or Higgs-strahlung off a vector boson
impossible. The analogous result for $\textrm{BR}(h \rightarrow
\tau^{+} \tau^{-})$ is plotted in Figure \ref{fig:Plot4}, with
similar conclusions.

\section{Phenomenology}
\label{sec:Phenomenology}

In this section we discuss the possibility of detecting a
supersymmetric leptophilic Higgs. We have focused on the neutral
sector, as the charged sector strongly resembles the non-SUSY
leptophilic scenario covered in \cite{Logan:2009uf}. The quantity of
importance to the decay of the lightest neutral scalar is the ratio
$\kappa = \textrm{BR}(h \rightarrow \tau^{+}\tau^{-})/\textrm{BR}(h
\rightarrow b\overline{b})$, which is given by equation
\ref{eq:Kappa BRtau-BRb Ratio} in Section \ref{sec:LEP Data}.

\begin{figure}[h]
\includegraphics[scale=.9]{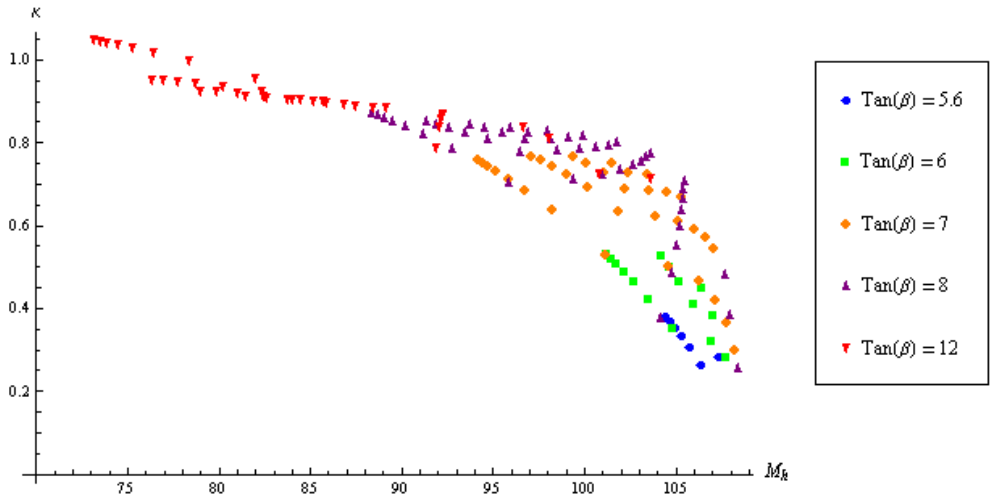}
\caption{ \small Various values of $\kappa$ plotted as a function of
the lightest neutral scalar mass $M_{h}$.} \label{fig:Plot5}
\end{figure}

For the region of parameter space discussed in the previous section,
we have shown various values of $\kappa$ in Figure \ref{fig:Plot5}.
For Higgs bosons near $114.4$ GeV, the allowed value of $\kappa$
approaches its Standard Model value of approximately $0.1$. However,
for lighter Higgs bosons, $\kappa$ is much bigger, approaching unity
for Higgs masses below $100$ GeV.

\begin{figure}[h]
\includegraphics[scale=1.05]{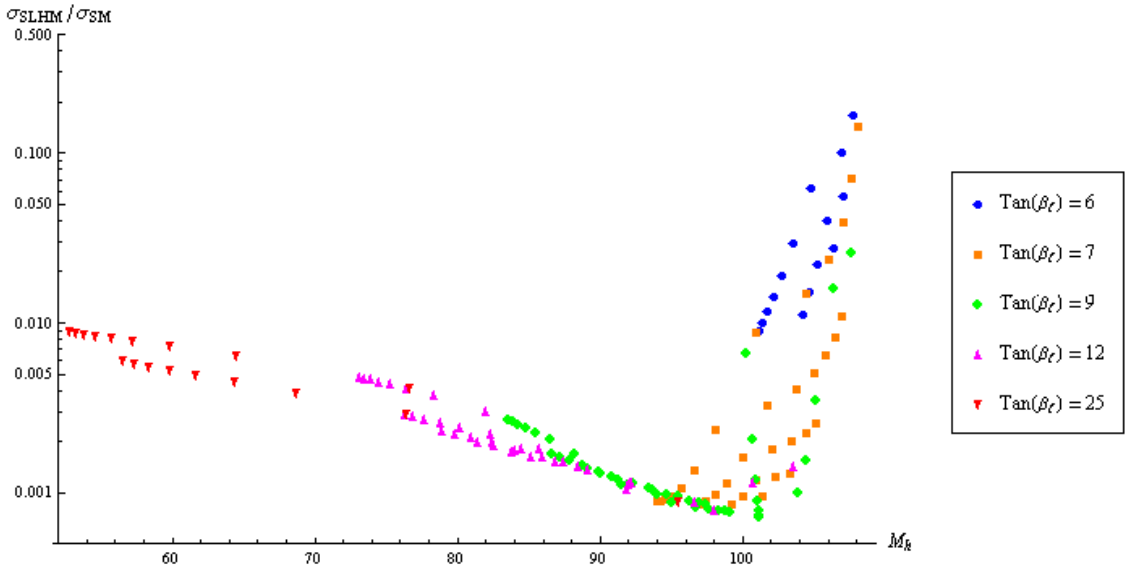}
\caption{ \small Logplot of the ratio of the production cross
section of the lightest neutral scalar by gluon fusion in the SLHM
to the Standard Model.} \label{fig:Plot6}
\end{figure}

We see that in this model, the Higgs can be relatively light, and
will have a much larger branching ratio to $\tau^+\tau^-$ than in
the Standard Model. In order to detect the Higgs at the Tevatron or
the LHC, however, one also must consider the production rate. As we
have seen, for Higgs bosons below $90$ GeV, the $ZZh$ and $WWh$
couplings are quite small, and thus Higgs-strahlung is negligible.
What about gluon fusion, which is the primary production mechanism
for a light Higgs? Here, one must include both top and bottom loops,
and the coupling to the Higgs will be different. We find that the
ratio of the gluon fusion cross section to that of the Standard
Model is \begin{equation} \label{eq:Gluon Fusion h Production}
\frac{\sigma_{SLHM}}{\sigma_{SM}} = \left|\frac{U_{11}}{\sin\alpha
\sin\beta} + \frac{A(m_{b})}{A(m_{t})} \frac{U_{21}}{\sin\alpha
\cos\beta}\right|^{2}, \end{equation} and this is plotted in Figure
\ref{fig:Plot6} for various parameters. The function $A(m_{f})$ is
given by $A(m_{f}) = 2\big[x_{f} + (x_{f} - 1)f(x_{f})
\big]x_{f}^{-2}$, where $x_{f} = M_{h}^{2}/4m_{f}^{2}$ and
$f(x_{f})$ is given by equation $2.47$ in \cite{Djouadi:2005gi}. For
much of parameter space, the gluon fusion rate is also very small,
making Higgs detection extremely difficult. In the Standard Model,
the only other production mechanism that doesn't involve gluon
fusion or the $WWh$ or $ZZh$ vertex is Higgs-strahlung off a top
quark. That is difficult in the Standard Model, and in this model is
even weaker since the top quark Yukawa coupling is smaller. One can
think about Higgs-strahlung off a tau, but this is likely to be
swamped by backgrounds.

In any event, this is just a specific model. One might have other
possibilities for Higgs production, such as production in the decay
of one of the charged Higgs bosons in the model, or production
through supersymmetric particles. In both of these scenarios, the
production rate would depend on many additional parameters. Thus,
experimenters should look for Higgs bosons in the $75-110$ GeV range
with a substantially enhanced coupling to $\tau$ pairs (below $75$
GeV, a very small sliver of parameter space does remain). A study of
$\tau$ pair detection in leptophilic Higgs decays at the LHC was
carried out in Ref. \cite{Belyaev:2009zd}. Since they did not
consider the supersymmetric version, they concentrated on Higgs in
the $100-160$ GeV mass range, and gluon fusion production was not
particularly suppressed, as it is here. They also focussed on models
with dark matter candidates (usually involving an additional singlet
or an additional inert doublet). Nonetheless, their techniques show
that detection of a Higgs decay into $\tau$ pairs is feasible in the
early stages at the LHC. At the Tevatron, CDF and D0 did explicitly
search for Higgs decays to $\tau$ pairs \cite{Kravchenko:2007qg},
but did not consider Higgs masses below $90$ GeV

Throughout this analysis, we have ignored the effects of the heavier
neutral Higgs scalars. Consider the second lightest neutral scalar,
$\eta$. As we scan the entire allowed parameter space, we find that
the $\eta$ always appears to be very close to 110 GeV. This may not
be too surprising. Imagine that there was no mixing at all between
the quarkophilic and leptophilic Higgs sectors. Then each sector
would have a similar mass matrix to that of the MSSM (although with
smaller overall vevs), and thus one would find two relatively light
Higgs. Mixing can't be eliminated, of course, due to D-terms, but it
is not surprising that there are two relatively light scalars in the
model. In the region of parameter space in which the couplings of
the $h$ to the gauge bosons is severely suppressed, however, the
couplings of the $\eta$ will not be, and thus the $\eta$ will be
similar to the Standard Model Higgs. Given the uncertainty in our
calculations, including the effects of non-leading-log and higher
order corrections to the masses, it is premature to conclude that
the current LEP bounds would rule out this $110$ GeV Higgs, but an
increase of just a few GeV in the current lower bound on the
Standard Model Higgs would rule out this model.

In the  region of parameter space of interest, the $h$ and $\eta$
are primarily linear combinations of $H_0$ and $H_u$, with small
admixtures of $H_d$ and $H_\ell$.   Nonetheless, the ratios of
vacuum expectation values are large enough that the dominant decay
of the $h$, for example, is primarily into $\tau$'s and $b$'s
through these small admixtures.    The two heaviest Higgs bosons are
each almost entirely $H_d$ and $H_\ell$ , respectively, with little
mixing.

Consider these two heavier Higgs bosons, $H_1$ and $H_2$.    Since
the coupling of the $\eta$, in the region of interest, to $Z$-pairs
is very close to that of the Standard Model, then the fact that the
sum of the squares of the Higgs couplings to $Z$-pairs must equal
the square of the Standard Model coupling implies that the coupling
of $H_1$ and $H_2$ with $W,Z$-pairs is negligible. We have confirmed
this numerically. Another way to say this is that the narrow window
of parameter space forces the direction of the vacuum expectation
value to be almost entirely in the $\eta$ direction, leaving little
room for vev-dependent couplings of the other neutral Higgs.   This
will also cause a suppression in the $H_1hh$ and $H_2hh$ couplings.
The $H_1$ and $H_2$ will thus be both Higgs-phobic and gauge-phobic
and will only decay into fermion pairs. One of the two, $H_1$, will
decay almost entirely into $b\bar{b}$, and the other, $H_2$, will
decay almost entirely into $\tau^+\tau^-$.    This leads to
interesting phenomenological consequences. The $H_1$ can be
copiously produced through gluon fusion (through its coupling to the
$b$-quark), and its dominant decay into $b\bar{b}$ will be quite
dramatic.    The $H_2$ would be a heavy Higgs boson that decays
entirely into $\tau$ pairs.   However, gluon fusion occurs at a
small rate, and thus production through heavier particles or
supersymmetric partners would be necessary.  This possibility is
currently under investigation.

\section{Conclusion}
\label{sec:Conclusion}

In this work, we have studied the Higgs sector of the supersymmetric
version of leptophilic models. The model contains four Higgs
doublets, which couple to the up quarks, down quarks, charged
leptons and no fermions, respectively. The Higgs sector, as in all
supersymmetric models, is tightly constrained. We consider
constraints from perturbativity, unitarity, the muon anomalous
magnetic moment and we also impose constraints from experimental
searches at LEP.

We find that in most of parameter space, the lightest Higgs, $h$,
has a mass between $75$ and $110$ GeV (with a very small sliver of
parameter space giving smaller masses). For lighter values of the
mass, the decay branching ratio into $\tau$ pairs is substantial,
and can even be the dominant decay mode. This would lead to some
spectacular signatures at the Tevatron and the LHC. However, the
conventional production mechanisms, such as W-fusion,
Higgs-strahlung and gluon fusion are suppressed in this region of
parameter space.

The second lightest Higgs, $\eta$, has a mass throughout the allowed
parameter space of approximately $110$ GeV. Its production cross
section is not as strongly suppressed, and would appear similar to a
Standard Model Higgs. The remaining two neutral scalars are
typically heavier, are gauge-phobic and Higgs-phobic, and would decay into fermions.    One decays almost entirely into $b\bar{b}$ and would be copiously produced through gluon fusion.  The other decays almost entirely into $\tau^+\tau^-$, but conventional production mechanisms are suppressed.

There are also three charged scalars and three pseudoscalars in the
model. We do not expect the phenomenology to differ substantially
from the detailed analysis of Logan and
MacLennan\cite{Logan:2009uf}, who used MSSM parameters to constrain
their parameter space (even though the model was not
supersymmetric), and thus there would only be $\mathcal{O}(1)$
changes in their results due to mixing angles. Exploration of the
supersymmetric particles in the model are currently under
investigation.

We thank Heather Logan and Reinard Primulando for useful discussions. This work was
supported by the National Science Foundation PHY-0755262.


\newpage
\begin{thebibliography}{99}

\bibitem{hhg}
  J.~F.~Gunion, H.~E.~Haber, G.~L.~Kane and S.~Dawson,
  ``THE Higgs Hunter's Guide,''  Addison-Wesley, Reading, USA, 1995.

\bibitem{Sher:1988mj}
  M.~Sher,
  Phys.\ Rept.\  {\bf 179}, 273 (1989).

\bibitem{Lee:1973iz}
  T.~D.~Lee,
  Phys.\ Rev.\  D {\bf 8}, 1226 (1973).

\bibitem{Weinberg:1976hu}
  S.~Weinberg,
  Phys.\ Rev.\ Lett.\  {\bf 37}, 657 (1976).

\bibitem{Branco:1985aq}
  G.~C.~Branco and M.~N.~Rebelo,
  Phys.\ Lett.\  B {\bf 160}, 117 (1985).

\bibitem{Liu:1987ng}
  J.~Liu and L.~Wolfenstein,
  Nucl.\ Phys.\  B {\bf 289}, 1 (1987).

\bibitem{Weinberg:1990me}
  S.~Weinberg,
  Phys.\ Rev.\  D {\bf 42}, 860 (1990).

\bibitem{Wu:1994ja}
  Y.~L.~Wu and L.~Wolfenstein,
  Phys.\ Rev.\ Lett.\  {\bf 73}, 1762 (1994)
  [arXiv:hep-ph/9409421].

\bibitem{Accomando:2006ga}
  E.~Accomando {\it et al.},
  arXiv:hep-ph/0608079.

\bibitem{Riotto:1999yt}
  A.~Riotto and M.~Trodden,
  Ann.\ Rev.\ Nucl.\ Part.\ Sci.\  {\bf 49}, 35 (1999)
  [arXiv:hep-ph/9901362].

\bibitem{Dine:2003ax}
  M.~Dine and A.~Kusenko,
  Rev.\ Mod.\ Phys.\  {\bf 76}, 1 (2004)
  [arXiv:hep-ph/0303065].

\bibitem{Peccei:1977hh}
  R.~D.~Peccei and H.~R.~Quinn,
  Phys.\ Rev.\ Lett.\  {\bf 38}, 1440 (1977).

\bibitem{Glashow:1976nt}
  S.~L.~Glashow and S.~Weinberg,
  Phys.\ Rev.\  D {\bf 15} (1977) 1958;   E.~A.~Paschos,
  Phys.\ Rev.\  D {\bf 15} (1977) 1966.

\bibitem{Su:2009fz}
  S.~Su and B.~Thomas,
  Phys.\ Rev.\  D {\bf 79} (2009) 095014
  [arXiv:0903.0667 [hep-ph]].

\bibitem{Logan:2009uf}
  H.~E.~Logan and D.~MacLennan,
  Phys.\ Rev.\  D {\bf 79} (2009) 115022
  [arXiv:0903.2246 [hep-ph]].

\bibitem{Goh:2009wg}
  H.~S.~Goh, L.~J.~Hall and P.~Kumar,
  JHEP {\bf 0905} (2009) 097
  [arXiv:0902.0814 [hep-ph]].

\bibitem{Cao:2009as}
  J.~Cao, P.~Wan, L.~Wu and J.~M.~Yang,
  Phys.\ Rev.\  D {\bf 80} (2009) 071701
  [arXiv:0909.5148 [hep-ph]].

\bibitem{Aoki:2009ha}
  M.~Aoki, S.~Kanemura, K.~Tsumura and K.~Yagyu,
  Phys.\ Rev.\  D {\bf 80}, 015017 (2009)
  [arXiv:0902.4665 [hep-ph]].

\bibitem{Logan:2010ag}
  H.~E.~Logan and D.~MacLennan,
  Phys.\ Rev.\ D {\bf 81}, 075016 (2010)
  [arXiv:1002.4916 [hep-ph]].

\bibitem{Davidson:2009ha}
  S.~M.~Davidson and H.~E.~Logan,
  Phys.\ Rev.\  D {\bf 80} (2009) 095008
  [arXiv:0906.3335 [hep-ph]].

\bibitem{Marshall:2009bk}
  G.~Marshall, M.~McCaskey and M.~Sher,
  Phys.\ Rev.\  D {\bf 81}, 053006 (2010)
  [arXiv:0912.1599 [hep-ph]].

\bibitem{Gunion:2002zf}
  J.~F.~Gunion and H.~E.~Haber,
  Phys.\ Rev.\  D {\bf 67}, 075019 (2003)
  [arXiv:hep-ph/0207010].

\bibitem{Gupta:2009wn}
  R.~S.~Gupta and J.~D.~Wells,
  Phys.\ Rev.\  D {\bf 81}, 055012 (2010)
  [arXiv:0912.0267 [hep-ph]].

\bibitem{Wells:2009kq}
  J.~D.~Wells,
  arXiv:0909.4541 [hep-ph].
\bibitem{Djouadi:2005gj}
  A.~Djouadi,
  Phys.\ Rept.\  {\bf 459}, 1 (2008)
  [arXiv:hep-ph/0503173].

\bibitem{Dawson:2010jx}
  S.~Dawson and P.~Jaiswal,
  arXiv:1009.1099 [hep-ph].

\bibitem{Jegerlehner:2007xe}
  F.~Jegerlehner,
  Acta Phys.\ Polon.\  B {\bf 38}, 3021 (2007)
  [arXiv:hep-ph/0703125].

\bibitem{Bennett:2004pv}
  G.~W.~Bennett {\it et al.}  [Muon g-2 Collaboration],
  Phys.\ Rev.\ Lett.\  {\bf 92}, 161802 (2004)
  [arXiv:hep-ex/0401008].

\bibitem{Bennett:2006fi}
  G.~W.~Bennett {\it et al.}  [Muon G-2 Collaboration],
  Phys.\ Rev.\  D {\bf 73}, 072003 (2006)
  [arXiv:hep-ex/0602035].

\bibitem{Barr:1990vd}
  S.~M.~Barr and A.~Zee,
  Phys.\ Rev.\ Lett.\  {\bf 65}, 21 (1990)
  [Erratum-ibid.\  {\bf 65}, 2920 (1990)].

\bibitem{Cheung:2003pw}
  K.~Cheung and O.~C.~W.~Kong,
  Phys.\ Rev.\  D {\bf 68}, 053003 (2003)
  [arXiv:hep-ph/0302111].

\bibitem{El Kaffas:2006nt}
  A.~W.~El Kaffas, W.~Khater, O.~M.~Ogreid and P.~Osland,
  Nucl.\ Phys.\  B {\bf 775}, 45 (2007)
  [arXiv:hep-ph/0605142].

\bibitem{WahabElKaffas:2007xd}
  A.~Wahab El Kaffas, P.~Osland and O.~M.~Ogreid,
  Phys.\ Rev.\  D {\bf 76}, 095001 (2007)
  [arXiv:0706.2997 [hep-ph]].

\bibitem{Amsler:2008zzb}
  C.~Amsler {\it et al.}  [Particle Data Group],
  Phys.\ Lett.\  B {\bf 667}, 1 (2008).

\bibitem{TeixeiraDias:2008fc}
  P.~Teixeira-Dias,p-
  J.\ Phys.\ Conf.\ Ser.\  {\bf 110}, 042030 (2008)
  [arXiv:0804.4146 [hep-ex]].

\bibitem{Barate:2003sz}
  R.~Barate {\it et al.}  [LEP Working Group for Higgs boson searches and
                  ALEPH Collaboration and  and],
  Phys.\ Lett.\  B {\bf 565}, 61 (2003)
  [arXiv:hep-ex/0306033].

\bibitem{Schael:2006cr}
  S.~Schael {\it et al.}  [ALEPH Collaboration and DELPHI Collaboration and
                  L3 Collaboration and ],
  Eur.\ Phys.\ J.\  C {\bf 47}, 547 (2006)
  [arXiv:hep-ex/0602042].

\bibitem{Achard:2003ty}
  P.~Achard {\it et al.}  [L3 Collaboration],
  Phys.\ Lett.\  B {\bf 583}, 14 (2004)
  [arXiv:hep-ex/0402003].

\bibitem{Abdallah:2004wy}
  J.~Abdallah {\it et al.}  [DELPHI Collaboration],
  Eur.\ Phys.\ J.\  C {\bf 38}, 1 (2004)
  [arXiv:hep-ex/0410017].

\bibitem{Djouadi:2005gi}
  A.~Djouadi,
  Phys.\ Rept.\  {\bf 457}, 1 (2008)
  [arXiv:hep-ph/0503172].

\bibitem{Belyaev:2009zd}
  A.~Belyaev, R.~Guedes, S.~Moretti and R.~Santos,
  JHEP {\bf 1007}, 051 (2010)
  [arXiv:0912.2620 [hep-ph]].

\bibitem{Kravchenko:2007qg}
  I.~Kravchenko  [CDF and D0 Collaboration],
in {\it In the Proceedings of the 15th International Conference on Supersymmetry and the Unification of Fundamental Interactions (SUSY07),
Karlsruhe, Germany, 26 Jul - 1 Aug 2007}
  arXiv:0710.5141 [hep-ex].


\end{thebibliography}
\end{document}